\documentclass[11pt,floatfix]{revtex4}
\usepackage{graphicx}
\usepackage{bm}
\usepackage{amsmath}
\usepackage{dcolumn}
\usepackage{bookman}

\begin{document}

\title{Density functional approach to study structural properties and Electric Field
Gradients in rare earth materials}

\author{S. Jalali Asadabadi}
\email[Electronic address: ]{sjalali@phys.ui.ac.ir}
\affiliation{Department
of Physics, Faculty of Science, University of Isfahan (UI)\\
Hezar Gerib Avenue, Isfahan 81744, Iran} \affiliation{Research
Center for Nano Sciences and Nano Technology\\ University of
Isfahan (UI), Isfahan 81744, Iran}
\author{H. Akbarzadeh}
\email[Electronic address: ]{akbarzad@cc.iut.ac.ir}
\affiliation{Department of Physics \\ Isfahan University of Technology \\ Isfahan, Iran}%

\date{\today}

\begin{abstract}
We investigated the effect of spin polarization on the structural
properties and gradient of electric field (EFG) on Sn, In, and Cd
impurity in RSn$_3$ (R=Sm, Eu, Gd) and RIn$_3$ (R=Tm, Yb, Lu)
compounds. The calculations were performed self-consistently
using the scalar-relativistic full potential linearized augmented
plane wave method. The local density approximations (LDA) and
generalized gradient approximation without spin polarization
(GGA) and with spin polarization (GGA+SP) to density functional
theory were applied. In addition to that we performed some
calculations within open core treatment ( GGA+open core). It is
clearly seen that GGA+SP is successful in predicting the larger
lattice parameter and the dramatic drop of EFG for R=(Eu, Yb)
relative to other rare earth compounds. This is an indication
that spin splitting generated by spin polarization without any
modification, is capable of treating properly the highly
correlated f electrons in these systems.
\end{abstract}

\maketitle

\section{\label{sec:level1}Introduction}
The highly correlated f electrons in rare earth materials have
challenged the reliability of the results obtained by
conventional density functional theory (DFT). It is generally
believed that regular DFT in its LDA/GGA formulation, without any
modification, is insufficient to correctly deal with the f
electrons; hence our understanding of the physical properties of
these systems are often based on a many body model Hamiltonian or
modified LDA such as LDA+U \cite{/NovP01/,/PriD00/,/AniV97/},
LDA+SIC \cite{/SvaA00/,/StrP99/}, and LDA+(open core)
\cite{/RicM98/} approaches.

Strange et.al \cite{/StrP99/}, by removing the spurious
self-interaction of f electrons reported a systematic
investigation of the rare earth elements and their sulfides. They
confirmed the pre-claimed idea that there are two types of f
electrons in these materials: Localized core-like f electrons
that determine the valency, and delocalized band-like f electrons
that are formed through hybridization with the s-d bands and
which participate in bonding. The main feature of their
calculation was based on an f band splitting into two sub-bands
one occupied, and the other unoccupied.

In a recent publication we investigated the valency of rare
earths in RIn$_3$; (R=Sm, Eu, Gd) and RSn$_3$; (R=Tm, Yb, Lu) and
calculated the electric field gradient using DFT without spin
polarization \cite{/JalS02/}. We found out that in regular DFT as
all f electrons are treated as band electrons, consequently their
density of states are all located in a single peak at the fermi
level which is not consistent with reality. We finally overcame
this drawback by applying open core approach.

The aim of the present work is to investigate the effect of spin
polarization on structural properties and the gradient of electric
field of the same samples. Our objective is to investigate the
borderline of LDA ability, that without any modification further
than spin polarization can obtain reliable results for some
compounds with highly correlated f electrons. Our calculations
confirm the fact that in most cases spin polarization, without
using LDA+U, LDA+SIC, or even open core modification can treat
the highly correlated f electrons properly. It means that at low
temperature, where the system is magnetically ordered, spin
polarization, by providing more flexibility among f electrons and
opening degrees of freedom for spin-splitting, partly compensate
the necessary coulomb correlation.

The organization of this paper is as follow. In Section II we
present the calculational details. Section III deals with the
results and discussions. Finally, we conclude this work with a
summary in Section IV.
\section{CALCULATIONAL DETAILS}
We utilize the full potential linearized augmented plane wave
(FP-LAPW) method within density functional theory. In this method
wavefunctions, charge density, and potential are expanded in
spherical harmonics within nonoverlapping muffin-tin spheres and
in plane waves in the remaining interstitial region of the unit
cell . As in FP-LAPW method no shape approximation is introduced
for potential and charge density, then it is an appropriate
procedure to obtain reliable results for the electric field
gradient (EFG) with high sensitivity to unisotropic charge
density distribution around the nucleus.

The EFG tensor is defined as the second derivative of the coulomb
potential with respect to the Cartesian coordinate at the nucleus,
which is written as a traceless tensor. The coulomb potential
which is one of the crucial quantities of an accurate
(full-potential) band structure calculation is given in LAPW as
lattice harmonics expansion inside the atomic spheres. In this
representation, the required derivatives can be determined
straightforwardly, and the largest component of EFG tensor  can
be obtained directly from the L=2, M=0 component of the potential
expansion inside the muffin tin spheres\cite{/BlaP88/}:
\begin{equation}
\label{eq-Vzz} V_{\mathit{zz}} = \sqrt{\frac{5}{4 \pi}}lim_{r\rightarrow 0}\frac{V_{20}}{r^2}
\end{equation}

It is assumed that the main axis of the EFG tensor points towards
z direction. The radial potential coefficient
$V_{20}\left(r\right)$ near the nucleus ($r\rightarrow 0$) for a
given charge density is obtained numerically by solving Poisson's
equation using a method proposed by Weinert \cite{/WeiM81/}.

The calculations were performed using the WIEN97 code developed
by Blaha and coworkers \cite{/BlaP99/}, which has been
successfully applied to a wide range of systems. In this code
core and valence states separated by their energy are treated
differently. Core states are treated within multiconfiguration
relativistic Dirac-Fock approach while valence states are only
scalar relativistic with an exception of $EuSn_3$ for which the
spin-orbit coupling in a second order variational steps were
included during self-consistent calculations of valence state. In
the later case the number of symmetry was reduced from 48 to 16
by assuming z-direction as a preferentional direction of the
coupled spin to the orbit. It was shown that by including the
spin-orbit interaction the results does not change substantially,
and as the incorporation of this coupling is very time consuming
then it was neglected throughout the calculation. For better
treatment of the core electrons we decomposed them into high
lying semicore and deeper true core states. As the electronic
charge of semicore states are not completely confined inside the
respective muffin-tin spheres, therefore we treated them as band
states in the full (nonspherical) potential. For these states in
addition to the usual LAPW basis we included local orbitals to
increase the flexibility of the basis set \cite{/SinD91/}, and
being able to use a sufficiently fine k sampling. This procedure
results the more accurate calculation of EFG, which is extremely
sensitive to the basis set \cite{/BomE89/}. The true core states
were treated atomic like in the self-consistent muffin-tin
potential. In our calculations electronic states with an energy
of at least 6 Ry below the Fermi energy were considered as true
core states. For the muffin-tin radius of rare earth elements as
well as Sn and In a value of 2.65 a.u. were chosen throughout
while for the Cd impurity manipulated into the RSn$_3$/In$_3$
compounds a radius of 2.5 a.u. was used. The maximum angular
momentum quantum number l as a cutoff for expanding the Kohn-Sham
wavefunctions in terms of lattice harmonics inside the muffin-tin
spheres was confined to $l_{max}=12$. The potential and charge
density was expanded inside the atomic spheres in crystal
harmonics up to L=6. The wavefunctions in the interstitial region
were expanded in plane waves with a cutoff of $K_{max}=7/R_{MT}$
where $R_{MT}$ is the smallest muffin-tin radius in the unit
cell. The periodic charge density and potential were Fourier
expanded up to $G_{max}=16$. We adjusted the value of mixing
parameter of charge density to 0.1 and used Broyden's scheme. We
have taken a mesh of 165 special k points to implement the valence
state integration in the irreducible wedge of the Brillioun zone
(BZ) that correspond to the grids $18\times18\times18$ in the
scheme of Monkhorst-Pack \cite{/MonH76/}. The mesh of k points,
charge density and potential expansion cutoff and finally
muffin-tin radii were allowed to vary for ensuring the
convergence. The exchange-correlation energy was calculated
within the local density approximation (LDA) based on the
Perdew-Zunger parameterization \cite{/PerJ81/},with and without
including advanced generalized gradient approximation (GGA) using
Perdew-Burke-Ernzerhof parameterization \cite{/PerJ96/}, with and
without considering spin polarization. For spin polarized
calculations , a ferromagnetic moment was imposed on R at the
start of the selfconsistency cycle. As for some compounds this is
not the true magnetic order \cite{/SanJ76/}, such an assumtion
should be considered as an extra possible source of error. At the
end of selfconsistency cycle, $YbIn_3$ and $LuIn_3$ came out to
be nonmagnet, all others remained ferromagnetic. In order to
calculate the electric field gradient at Cd site as a
representation of an impurity atom in each compound all sides of
the unit cell were doubled to construct a supercell with 16 atoms
\cite{/JalS00/}. Further increase of the sides of supercell did
not change the results significantly. To keep the same accuracy
with the previous primitive unit cell the mesh of k points in
supercell was reduced to 75 corresponding to the
$9\times9\times9$ grids. The energy for each orbital in $EuSn_3$,
and in $YbIn_3$ is linearized by selecting the most probable
energy calculated for each density of state over the valence and
semicore energy windows. The linearizations obtained for $EuSn_3$
and $YbIn_3$ then are used as an initial input for the $SmSn_3$,
$GdSn_3$, and the $TmIn_3$, $LuIn_3$, respectively and we allow
the program to search automatically for the best-linearized
energies of Cd orbitals and check if they are perfect.

\section{RESULTS AND DISUSION}
\subsection{\label{sec:level2}Structural Properties}
By calculating the total energy of a primitive unit cell as a
function of its volume and fitting the data with the Birch
equation of state \cite{/BirF78/} we obtained the   lattice
parameters, bulk moduli, and the pressure derivative of the bulk
moduli for RSn$_3$ (R=Sm, Eu, Gd) and RIn$_3$ (R=Tm, Yb, Lu) by
LDA, GGA, GGA+SP, GGA +open core (valence 2, 2.5 and 3)
approaches. These results along with the experimental data are
listed in Tables \ref{table1} and \ref{table2}.

\begin{table}
\caption{\label{table1}The comparison of the measured lattice
parameters in $\AA$ at T=300K with the theoretical values
calculated at T=0K within LDA, GGA without spin polarization
(GGA), GGA with spin polarization (GGA+SP), and GGA+open core
(valence 2, 2.5, 3).}
\begin{ruledtabular}
\begin{tabular}{ccccccc}
 &SmSn$_3$&EuSn$_3$&GdSn$_3$&TmIn$_3$&YbIn$_3$&LuIn$_3$\\ \hline
 LDA&$4.529$&$4.515$&$4.513$&$4.434$&$4.443$&$4.434$ \\
 GGA&$4.654$&$4.649$&$4.648$&$4.589$&$4.594$&$4.566$ \\
 GGA+SP&$4.708$&$4.715$&$4.683$&$4.593$&$4.594$&$4.566$ \\
 $GGA^{divalent}$&$4.809$&$4.784$&$4.749$&$4.651$&$4.626$&$4.655$ \\
 $GGA^{di~\&halfvalent}$&$-$&$4.758$&$4.733$&$4.619$&$4.604$&$-$ \\
 $GGA^{trivalent}$&$4.747$&$4.732$&$4.717$&$4.587$&$4.582$&$4.576$ \\
 \footnote{{Ref.}{\cite{/SchG77/,/SanJ76/}}}Expt.&$4.69$&$4.75$&$4.68$&$4.56$&$4.61$&$4.55$ \\
\end{tabular}
\end{ruledtabular}
\end{table}

The lattice parameters calculated by GGA for all compounds are in
better agreement with the experiment than those obtained by LDA.
The better prediction of experimental results by GGA has also been
observed in most rare earth metals and is believed to be due to
the fact that the nonlocality of exchange and correlation is
better taken into account by GGA than LDA \cite{/DelA98/}. We
expect the lattice constant for divalent Eu and Yb to be higher
than the other rare earth compounds. The divalent character
results to weaker interatomic bonds and manifests itself in a
higher lattice parameter and also compressibility, which is the
inverse of bulk modulus. It is clearly seen that GGA fails to
predict such behavior for Eu compound, while GGA+SP improves the
results in the right direction. Richter \cite{/RicM98/} by
comparing the charge densities belonging to non-magnetic and
polarized solutions of 4f electrons of gadolinium atom has
illustrated a clear reason why spin-independent DFT is not suited
to magnetic systems. Our results confirm his conclusions. As
expected, due to rather full f shell in Lu, Yb and even Tm
compounds, spin polarization does not practically affect their
lattice parameter. For $SmSn_3$ and $GdSn_3$ the results obtained
by GGA+SP are in nice agreement with experiment while for Eu and
Yb compounds the results obtained by open core calculation with a
mixed valence ($\sim2.5$) are in better agreement with
experiment. A point worth to mention is that the deviation of
lattice parameter calculated by GGA+SP and GGA+open core are more
significant in Sm, Eu, and Gd compared to Tm, Yb, and Lu
compounds. It is probably because of the fact that going from
left to right in the list of lantanides the f electrons are
getting more localized. Hence, for the highly localized f
electrons in Tm, Yb, and Lu compounds it does not make any
difference if we leave them alone or confine them inside the core
by open core approach. Similar behavior has been predicted for f
electrons in actinides \cite{/PetL01/}. Finally, it should be
noticed that although the equilibrium lattice parameters can be
accurately extracted by fitting the total energy-volume data with
an appropriate equation of state, but the value obtained for the
bulk modulus and its pressure derivative are sensitive to the
range of fitting. Hence, for the results presented in Table
\ref{table2} only the qualitative analysis are reliable and we
can not fully trust the quantitative values. Furthermore as to
our knowledge no experimental data for the bulk modulus and its
derivative has been reported yet, hence our results serve only as
a prediction for future studies.

\begin{table}
\caption{\label{table2}The calculated Bulk modulus in GPa and its
pressure derivative within LDA, GGA without spin polarization
(GGA), and GGA with spin polarization (GGA+SP).}
\begin{ruledtabular}
\begin{tabular}{ccccccc}
 &SmSn$_3$&EuSn$_3$&GdSn$_3$&TmIn$_3$&YbIn$_3$&LuIn$_3$\\ \hline
 $B_0$(LDA)&$65.517$&$59.786$&$62.549$&$73.032$&$70.596$&$77.468$ \\
 $B^\prime_0$(LDA)&$2.509$&$2.312$&$2.948$&$5.789$&$5.089$&$4.708$ \\
 $B_0$(GGA)&$63.531$&$61.020$&$59.629$&$53.440$&$55.180$&$62.604$ \\
 $B^\prime_0$(GGA)&$3.478$&$3.401$&$3.560$&$4.937$&$4.486$&$4.159$ \\
 $B_0$(GGA+SP)&$58.011$&$54.239$&$64.318$&$55.788$&$55.081$&$62.648$ \\
 $B^\prime_0$(GGA+SP)&$4.354$&$5.064$&$4.688$&$4.709$&$4.402$&$4.155$ \\
\end{tabular}
\end{ruledtabular}
\end{table}

\subsection{\label{sec:level3}Electric Field Gradient}

In the literature, different conversion relations analyze the
experimental results of the nuclear quadrupole interaction. Then
to compare our calculated EFG with experiment we first need to
specify the conversion relation that has been used. The
experimental data to calculate the EFG on Sn site of RSn$_3$;
(R=Sm, Eu, Gd) compounds are obtained from Mossbauer
spectroscopy. Hence the value of $V_{zz}$ can be calculated from
the measured Doppler velocity $\Delta_\nu$ via :
\begin{equation}
\label{eq-mms} V_{\mathit{zz}}[10^{21}Vm^{-2}] = 12.79\Delta_\nu(mms^{-1})
\end{equation}
While the EFG on a Cd impurity substituted at Sn and In sites
were obtained by Perturbed Angular Correlation measurements with
the following conversion relation:

\begin{eqnarray}
V_{\mathit{zz}}[10^{21}Vm^{-2}]=\frac{0.041355}{Q(b)}\omega(MHz)
\end{eqnarray}
We used the recently proposed value of Q=0.83b for nuclear quadrupole moment of Cd.
\begin{table}
\caption{\label{table3}The comparison of the derived EFG in
10$^{21}$V/m$^2$ from the measured quadrupole interactions of
$^{119}$Sn(Q=0.124,$E_0$=23.8keV) at T=77K and 4.2K with the
theoretical values calculated at T=0K within LDA, GGA without
spin polarization (GGA), and GGA with spin polarization (GGA+SP).
}
\begin{ruledtabular}
\begin{tabular}{cccc}
 &SmSn$_3$&EuSn$_3$&GdSn$_3$\\ \hline
 LDA&$17.49$&$16.56$&$16.51$ \\
 GGA&$16.71$&$15.35$&$15.39$ \\
 GGA+SP&$15.20$&$14.03$&$15.20$ \\
 \footnote{{Ref.}{\cite{/SanJ76/}}}$Expt.^{T=4.2K}$&$13.82$&$13.56$&$-$ \\
 \footnotemark[1]$Expt.^{T=77K}$&$13.82$&$11.64$&$13.56$ \\
\end{tabular}
\end{ruledtabular}
\end{table}

\begin{figure*}
 \begin{center}
  \includegraphics[width=15cm,angle=0]{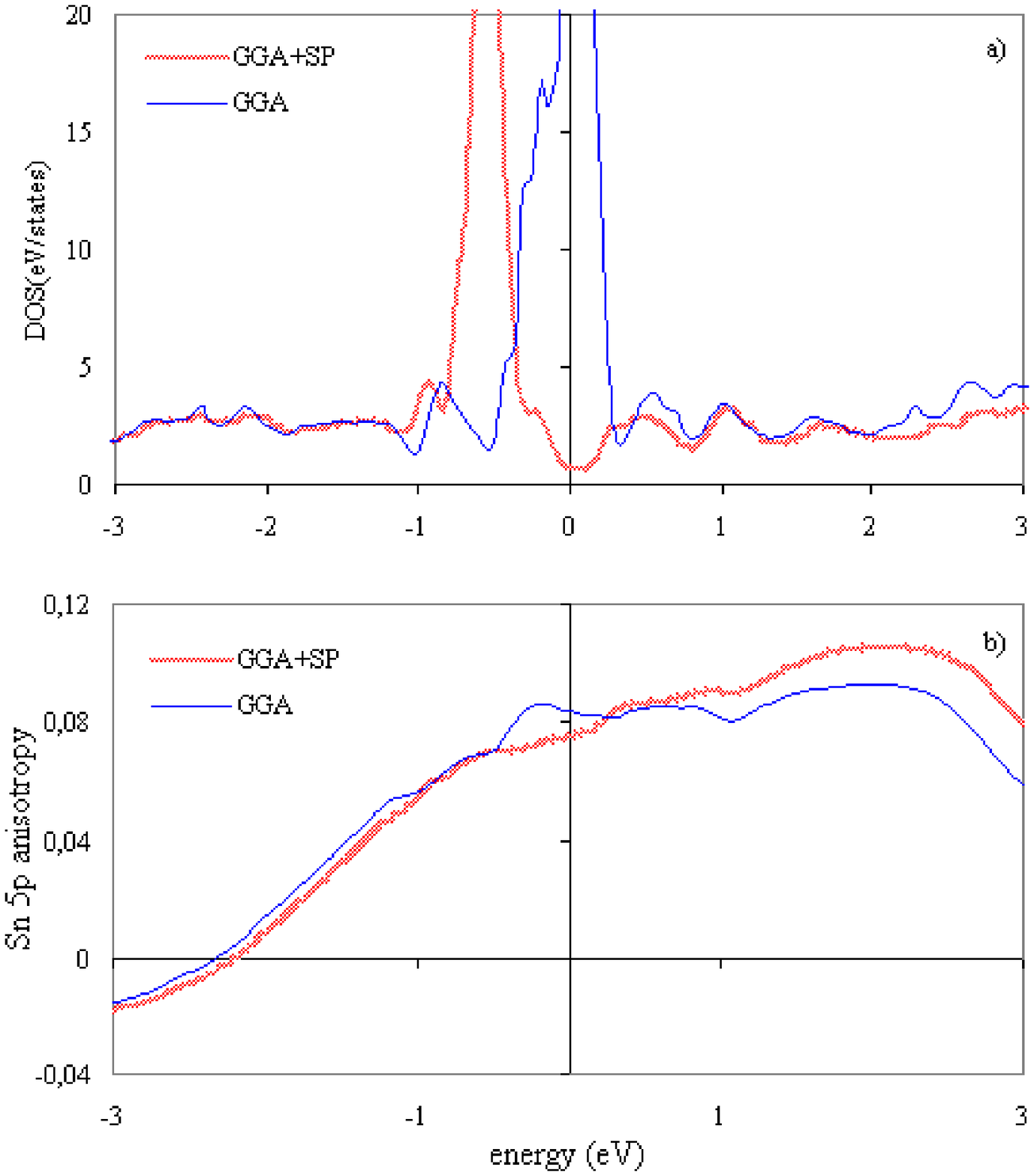}
   \caption{All graphs are for EuSn3. The solid line is for GGA, and faint line for GGA+SP.
   (a) total DOS, (b) Sn 5p anisotropy function.
    \label{fig1}}
 \end{center}
\end{figure*}

The calculated EFG on Sn sites of RSn$_3$; (R= Sm, Eu, Gd) are
compared with the experimental data in Table \ref{table3}. For
all compounds, the less agreement with experiment is due to LDA
calculations. GGA + SP correctly predict the lower value of EFG
in $EuSn_3$ compared to the other two compounds which is
consistent with the experimental results obtained at nonzero
temperatures. As EFG is very sensitive to the charge density
around the nucleus, then its nearly constant value at Sn sites
for all three compounds confirm the argument presented in Ref.
\cite{/SanJ76/} that the charge density around Sn nucleus are
rather identical and independent of rare earth atoms. In Table
\ref{table4} we compare the calculated EFG on Cd impurity with
experiment \cite{/SchG77/}. Here also, the results obtained by
GGA+SP are in better agreement with experiment. The EFG drop for
Eu and Yb compounds, attributed to a changing valency, is clearly
visible. The small deviation between calculation and experiment
presented in Tables \ref{table3} and \ref{table4} are not
unexpected given the uncertainties of the temperature effect,
which are not included in our calculation. It seems that the
electric field gradient at most cases decreases when the
temperature rises. The EFG measured in $YbIn_3$ and $EuSn_3$ at
two different temperatures given in Tables \ref{table3} and
\ref{table4} confirm this expectation. In addition to that, to
avoid time consuming, we have not relaxed the atomic position.
Since EFG may strongly be affected by local geometry, then the
relaxation of the atoms from their original lattice positions can
alter their value \cite{/LanS00/}, hence the atomic unrelaxation
can be considered as an another source of error in our
calculation. The last possible source of error is that we have
used experimental lattice parameter to calculate EFG which are
not identical with the value obtained from E-V curves. Although
we will show later that the effect of this discrepancy is
ignorable.

\begin{figure*}
 \begin{center}
  \includegraphics[width=19cm,angle=0]{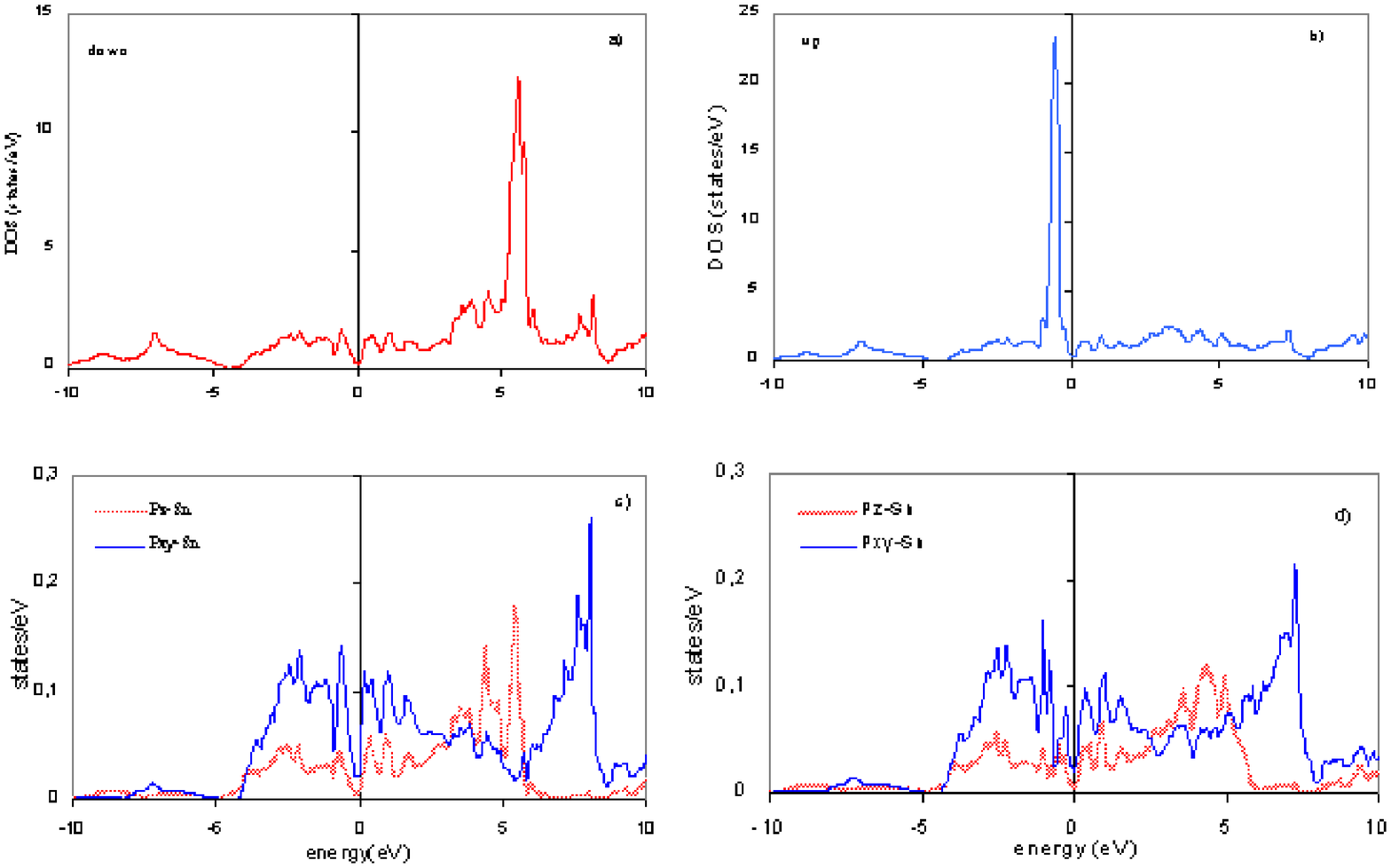}
   \caption{All graphs are for EuSn3. (a) total spin-up DOS, (b) total spin-down
   DOS. The changes of pz (faint curves) and pxy (solid curves) as a function of energy are
   shown in (c) for spin-up and in (d) for spin-down.
   \label{fig2}}
 \end{center}
\end{figure*}

\begin{table}
\caption{\label{table4}The comparison of the derived EFG in
10$^{21}$V/m$^2$ from the measured quadrupole coupling constants
of $^{111}$Cd(Q=0.83b, $E_0$=247keV) at T=300K and 4.2K with the
theoretical values calculated at T=0K within GGA without spin
polarization (GGA), and GGA with spin polarization (GGA+SP).}
\begin{ruledtabular}
\begin{tabular}{ccccccc}
 &SmSn$_3$&EuSn$_3$&GdSn$_3$&TmIn$_3$&YbIn$_3$&LuIn$_3$\\ \hline
 GGA&$4.06$&$3.44$&$3.64$&$3.36$&$3.11$&$6.08$ \\
 GGA+SP&$2.62$&$1.81$&$3.76$&$3.49$&$3.12$&$6.08$ \\
 \footnote{{Ref.}{\cite{/SchG77/}}}$Expt.^{T=300K}$&$2.12$&$0.63$&$2.12$&$4.37$&$1.92$&$4.32$ \\
 \footnotemark[1]$Expt.^{T=4.2K}$&$-$&$-$&$-$&$-$&$2.16$&$-$ \\
\end{tabular}
\end{ruledtabular}
\end{table}
At this stage we raise a question that our ab initio study is
going to answer. To which extent the EFG at X in a $RX_3$
compound is sensitive to the behavior of f electrons in R and how
the spin polarization can push the system in right direction? In
order to tackle this question, we will follow Ref.
\cite{/JalS02/} by introducing a visualization tool called
anisotropy function $\Delta p\left(E\right)$. This function
allows us a transparent interpretation of the qualitative
behavior of Vzz. As Vzz expresses the deviation from spherical
symmetry of the electron density in the environment of X, hence
this aspherically is given in terms of p-orbitals (s and d
orbitals have ignorable effects \cite{/JalS02/}) by following
expression:
\begin{eqnarray}
\label{eq-pdspher} V_{\mathit{zz}} & \approx & V_{\mathit{zz}}^p
\\ V_{\mathit{zz}}^p & = & \Delta p\left(E_F\right) \, \left<
\frac{1}{r^3} \right>_p
\\ \label{eq-p-aniso} & & \Delta p\left(E_F\right) = \frac{1}{2} p^I_{xy}\left(E_F\right)
- p^I_z\left(E_F\right) \\ \label{eq-pInt} & & p^I_i(E_1) =
\int_{-\infty}^{E_1} p_i(E) dE
\end{eqnarray}

Here $\left< \frac{1}{r^3} \right>_p$ is an expectation value for
the p-orbitals, $p_z\left(E\right)$ is the partial p-DOS (Density
of States) in the muffin tin sphere around an atom, and $E_F$ is
the Fermi energy. The integral $p^I_i\left(E_1\right)$ counts the
number of $p_i$ electrons in a muffin-tin sphere with energy less
than $E_1$.

In order to investigate the effect of spin polarization on the
value of EFG, the total DOS of EuSn3 within GGA and GGA+SP are
shown in Fig1.a. The only considerable effect of spin
polarization is the shift of f-peak from the initial fermi level
to the left. Due to the hybridization of f elections on R sites
with p electrons on neighboring X cites, the shift of f peak
affect Pxy and Pz and subsequently $\Delta p\left(E_F\right)$. The
changes of anisotropy function with energy are shown in Fig1.b.
It is clearly seen that $\Delta p\left(E_F\right)$ and as a
result of that EFG obtained by GGA +SP is lower than the
corresponding values within GGA.

To analyze the origin of the EFG on Cd, Sn, and In sites, we
decomposed the calculated EFG into several different
contributions (Table \ref{table5}). The valence EFG
(${}^{val}V_{zz}$) originates from the asphericity of the valence
(and semicore) charge density inside the muffin tin sphere. The
lattice EFG (${}^{lat}V_{zz}$) arises from the boundary value
problem and charge distribution outside the muffin-tin sphere.
${}^{val}V_{zz}$ is further decomposed into spin up
(${}^{val}V^\uparrow_{zz}$) and spin down
(${}^{val}V^\downarrow_{zz}$) contributions. It is seen that the
lattice contribution is negligibly small (less than 2\% of the
total EFG), which reflects the fact that the main origin of the
EFG lies in the deviation from spherical symmetry of the valence
charge density near the nucleus. It also confirm that the total
EFG is not highly sensitive to the lattice parameters, hence the
non-optimized experimental lattice constant used in our
calculation is acceptable.
${}^{val}V^\uparrow_{zz}$-${}^{val}V^\downarrow_{zz}$ is expected
to be a qualitative criteria for magnetism of the system; the
vanishing of this quality for non-magnetic compounds such as
YbIn$_3$ and LuIn$_3$ confirm this expectation.

\begin{figure*}
 \begin{center}
  \includegraphics[width=15cm,angle=0]{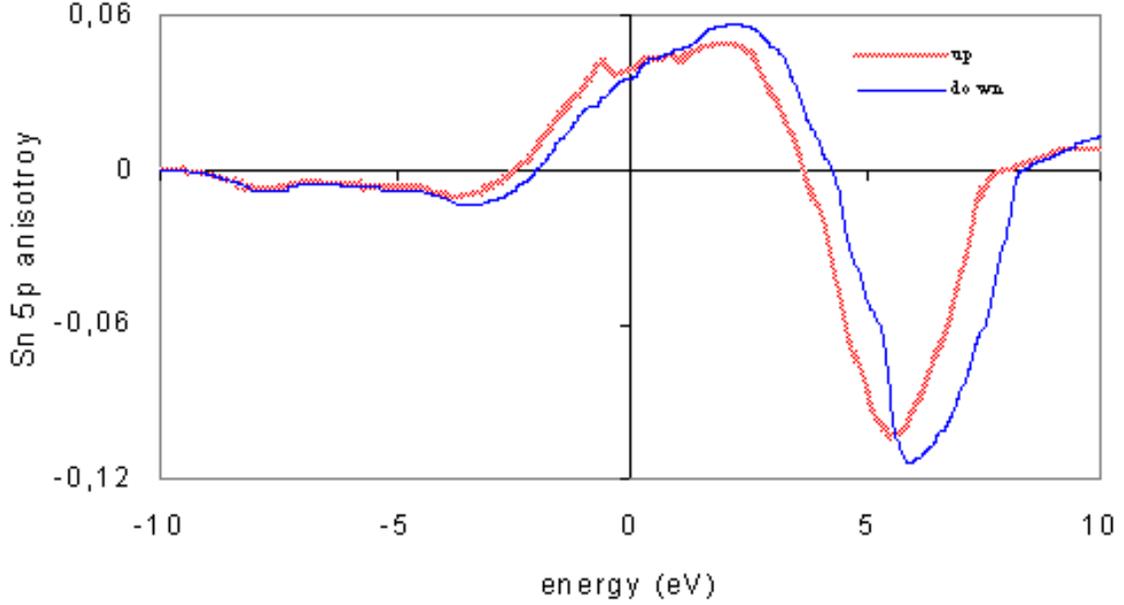}
   \caption{The changes of anisotropy function $\Delta p(E)$ for both up (faint) and down
   (solid) spin directions with respect to energy.
   \label{fig3}}
 \end{center}
\end{figure*}

\begin{table}
\caption{\label{table5}The decompositions of calculated EFG's at
Cd, Sn and In sites into the valence EFG for spin up and down and
the total lattice EFG in 10$^{21}$V/m$^2$.}
\begin{ruledtabular}
\begin{tabular}{ccccc}
&LuIn$_3$\\ \hline
 &$V^{latt}_{zz}$&${}^{val}$$V_{zz}$&${}^{val}$$V^\uparrow_{zz}$&${}^{val}$$V^\downarrow_{zz}$\\
 \hline
 &$$&$Cd$&$$& \\
 SmSn$_3$&$-0.03$&$2.62$&$2.26$&$0.39$ \\
 EuSn$_3$&$-0.03$&$1.81$&$1.49$&$0.35$ \\
 GdSn$_3$&$-0.04$&$3.76$&$2.18$&$1.62$ \\
 TmIn$_3$&$-0.03$&$3.49$&$1.70$&$1.82$ \\
 YbIn$_3$&$-0.00$&$3.12$&$1.56$&$1.56$ \\
 LuSn$_3$&$-0.02$&$6.08$&$3.05$&$3.05$ \\
 &$$&$Sn$&$$& \\
 SmSn$_3$&$-0.02$&$15.02$&$8.49$&$6.73$ \\
 EuSn$_3$&$-0.03$&$14.03$&$7.33$&$6.73$ \\
 GdSn$_3$&$-0.02$&$15.83$&$7.39$&$8.46$ \\
 &$$&$In$&$$& \\
 TmIn$_3$&$-0.01$&$8.95$&$4.22$&$4.74$ \\
 YbIn$_3$&$-0.01$&$8.34$&$4.18$&$4.18$ \\
 LuIn$_3$&$-0.04$&$11.23$&$5.62$&$5.62$ \\

\end{tabular}
\end{ruledtabular}
\end{table}

At the end to analyze the up and down contributions to Vzz on Cd,
Sn and In in the listed compound we concentrate, as an example,
on Sn site in $EuSn_3$. The spin up and down DOS and their
corresponding $p_z$ and $p_{xy}$ on Sn cites are shown in Fig2.
First of all, it is seen that spin polarization has splited the
up and down f peaks (Fig2.a, and b). This feature is compatible
with reality. In nature there are two types of f electrons,
localized and nonlocalized, and as a result of that the f band
should split into two sub-bands. In regular GGA as the strong
correlation between the f electrons are not treated properly,
then all of them are located in a single peak at $E_F$. Hence it
is not unjustified to expect that magnetism push the system in
the right direction. How the f-peak splitting affects $p_z$,
$p_{xy}$, $\Delta p(E_F)$, and consequently EFG? Fig2c and d show
the changes of $p_z$ and $p_{xy}$ with energy for up and down
spin separately. Below -4eV the two set of curves coincide. In
the region [-4, 0 ] eV $P^\uparrow_{xy}$ is slightly larger than
$P^\downarrow_{xy}$. This difference is more visible at around -1
eV, which is the location of $f^\uparrow$ peak. One can conclude
that the Sn $p^\uparrow$-electrons in the xy-plane hybridizes
with the f-electrons of the neighboring rare earths (which are
all 4 in the xy-plane). In Fig.3 the anisotropy function $\Delta
p(E)$ is given for two spin directions (call it $\Delta
p^\uparrow(E)$ and $\Delta p^\downarrow(E)$). At fermi level,
$\Delta p^\uparrow(E)$ is slightly larger than $\Delta
p^\downarrow(E)$. Exactly the same was seen for $V_{zz}$ in Table
\ref{table3}, which demonstrates once more that the p-anisotropy
really reflects the behavior of $V_{zz}$. As we already expected
from $p_z$ and $P_{xy}$ curves, no difference is observed between
the two curves below -4 eV. In the region [-4, 0 ] $\Delta
p^\uparrow(E)$ is larger than $\Delta p^\downarrow(E)$ and at
around -1 eV $p^\uparrow$ anisotropy shows a clear peak. This is
due to the $f^\uparrow$ peak located in this area.

\section{conclusion}
We employed the FP-LAPW method to calculate the structural
properties of RSn$_3$ ( R=Sm, Eu, Gd) and RIn$_3$ ( R=Tm, Yb, Lu
) by LDA, GGA, GGA+SP, and GGA+open core ( valence 2, 2.5, and 3
) approaches. For Eu and Yb compounds open core and for other
compounds GGA+SP results are in better agreement with experiment.
We also calculated the EFG at Sn, In, and impurity Cd sites in
RSn$_3$ (R=Sm, Eu, Gd) and RIn$_3$ (R=Tm, Yb, Lu) compounds. Here
also, the results obtained by GGA+SP are in better agreement with
experiment. The reasonable agreement of the results calculated by
GGA+SP with experiment confirm the fact that even for such highly
correlated systems the spin splitting introduced by magnetism can
compensate the strong coulomb correlation. Our approach has the
advantage of treating the localized and itinerant electronic
states on the same common footing.

\begin{acknowledgments}
S.J.A., and H.A. acknowledge the financial support of the Isfahan
University of Technology and also the Association Scheme of the
Abdus Salam International Centre for Theoretical Physics (ICTP).
\end{acknowledgments}

\end{document}